\documentstyle[prl,aps,preprint]{revtex}
\preprint{\vbox{\hbox{ACT-07/95}
\hbox{CERN-TH/95-88}\hbox{CTP-TAMU-21/95}
\hbox{UMN-TH-1349/95}\hbox{SNUTP 95-041}
\hbox{hep-ph/9505438}}}
\begin{document}
\title{
On the Thermal Regeneration Rate for Light Gravitinos\\
in the Early Universe}

\author{John Ellis${}^1$, D.V. Nanopoulos${}^{2,3}$,
Keith A. Olive${}^4$ and Soo-Jong Rey${}^5$}

\address{
Theory Division, CERN, CH-1211 Geneva SWITZERLAND${}^1$\\
Department of Physics, Texas A\&M University,
College Station, TX 77843 USA${}^2$\\
Astroparticle Physics Group,
Houston Advanced Research Center (HARC),\\
The Mitchell Campus, The Woodlands, TX 77381, USA${}^3$\\
Department of Physics, University of Minnesota,
Minneapolis MN 55455 USA${}^4$\\
Physics Department and Center for Theoretical Physics,
Seoul National University, Seoul 151-742 KOREA${}^5$}

\maketitle
\vskip -0.2in
\abstract
We investigate the light gravitino regeneration
rate in the early Universe in
models based on $N=1$ supergravity. Motivated by
a recent claim by Fischler,
we evaluate finite-temperature effects on the
gravitino regeneration rate due
to the hot primordial plasma for a wide range of
the supersymmetry-breaking
scale $F$. We find that thermal corrections to the
gravitino pole mass and to
the Goldstino coupling are negligible for a wide
range of temperatures, thereby
justifying the extension of the equivalence theorem
for the helicity-1/2
gravitino and Goldstino to a hot primordial plasma
background. Utilizing the Braaten-Pisarski resummation method,
the helicity-1/2 gravitino regeneration rate
is found to be
$ 0.25 \alpha_s(T) \log(1 / \alpha_s(T))
|{m_{\rm soft} / F}|^2 T^3(1 + \alpha_s(T)
\log(1/\alpha_s(T)) + T^2 / |F|)$
up to a calculable, model-dependent ${\cal O}(1)$
numerical factor. We review the implications of
this regeneration rate for supergravity cosmology,
focusing in particular on scenaria for baryogenesis.

\def\be{\begin{equation}}
\def\ee{\end{equation}}
\def\bee{\begin{eqnarray}}
\def\eee{\end{eqnarray}}
\def\SG{\Sigma_G(E, {\vec k})}
\def\ms{m_{\rm soft}}
\def\la{{{\lower 5pt\hbox{$<$}} \atop {\raise 5pt\hbox{$\sim$}}}~}
\def\ga{{{\lower 2pt\hbox{$>$}} \atop {\raise 1pt\hbox{$\sim$}}}~}

\section{Introduction}

One of the most serious cosmological
constraints in the framwework of $N=1$ supergravity
is the gravitino problem \cite{weinberg} \cite{eln}.
Whilst inflation dilutes
exponentially any pre-existing gravitinos,
thermal gravitinos may be regenerated in significant numbers
if the post-inflationary
reheating temperature is sufficiently high. Once regenerated,
thermal gravitinos may lead to a cosmological disaster. Long-lived
gravitinos may dominate the
energy density of the early Universe, or their decays may alter the
light-element abundances
calculated in primordial nucleosynthesis, if the gravitinos decay
after nucleosynthesis. Therefore, any successful supersymmetric
inflationary model should keep the reheating temperature low
enough that sufficiently few
gravitinos are regenerated subsequently. This requirement is known
to put a severe upper bound on the reheating temperature
\cite{other}-\cite{morth}, unless the gravitino is very light.

At energies that are sufficiently high
compared to the gravitino mass, it is known
that interactions of the helicity-1/2 components of the
gravitino dominate over the helicity-3/2 components.
Fischler \cite{fischler} recently re-analyzed the gravitino
regeneration rate and suggested that heat-bath effects might
enhance greatly this rate at high temperature.
Fischler's argument was based on the well-known fact
that supersymmetry (SUSY) is broken \cite{boyanovsky} at finite
temperature \cite{daskaku} \cite{girardello},
because the thermal distributions are different for
bosons and fermions. He argued that the gravitino regeneration rate
should be proportional to $T^5$,
instead of $T^3$ as estimated previously \cite{nos}.
If correct, his result would imply that the reheating temperature
should not exceed $\sim 10^4 - 10^7 $ GeV,
or the gravitino should be lighter than 1 KeV or heavier
than several TeV.
Such a conclusion would be
significant for both cosmology and supersymmetric phenomenology,
requiring either a low reheating temperature or an
extremely low/high gravitino mass,
with a correspondingly low/high supersymmetry breaking scale
$\sqrt {|F|} \sim 10^6/10^{11}$ GeV.
Motivated by these considerations, in this paper we study
finite-temperature effects on the gravitino regeneration rate
in a more systematic way.

At finite temperature there are various thermal effects one
has to take into account in order
to understand the chemical equilibration of light gravitinos.
First, the heat bath may generate temperature-dependent mass
renormalization.
Since the gravitinos are light, one might worry that the thermal
mass renormalization effect could be
significant. If the thermal mass renormalization is so large
that it exceeds the typical energy of the gravitino or the
Goldstino-matter coupling strength becomes strong, then one cannot
expect any more that the helicity-1/2 components of gravitinos are
dominant in regeneration processes. Secondly, the reaction rate
for chemical equilibration has to be calculated using the
full panoply of thermal field theory. The Landau damping
phenomena in the electron-gas plasma \cite{fetterwalecka}
and hot gauge theories are well-known examples.
Finite-temperature effects enter through
the effective thermal propagators, vertices and
through the external particles' thermal phase space factors.
In this paper,
we adopt the method proposed by
Braaten and Pisarski \cite{braatenpisarski} for taking
these effects correctly
into account, and calculating consistently thermal reaction rates.
In this method, the
underlying spontaneously-broken supersymmetry may not be manifest.
Not only does the heat bath distinguish bosons and fermions
through their different equilibrium distribution functions,
but also intrinsically finite-temperature effects such as
collective excitations are more significant for
fermions and gauge bosons than for
scalar particles. However, our explicit calculations show that
thermal effects are quite insignificant,
in that the finite-temperature corrections to the gravitino
mass and to the Goldstino-matter coupling are negligible, and the
zero-temperature estimates of the gravitino regeneration rate are
qualitatively correct. This conclusion
is in disagreement with Fischler's argument.

This paper is organized as follows.
In Section 2,  we review the
low-energy effective Lagrangian for the
helicity-3/2 and -1/2 components of a
light gravitino. Then we calculate temperature-dependent
corrections to the gravitino mass,
and show that they are proportional to $(T/ M_P)^2 \, \ms$,
where $\ms$ represents a typical soft supersymmetry-breaking
mass splitting for a chiral or gauge supermultiplet.
Similarly, we find a thermal correction to the Goldstino-matter
coupling due to thermal Goldstino dynamics that
is proportional to $T^2 / |F|$.
These corrections are numerically negligible for
the range of supersymmetry-breaking scale $\sqrt {|F|}$
of phenomenological interest. Therefore we may continue to
use the equivalence theorem for light helicity-1/2 gravitinos
and Goldstinos.
In Section 3, utilizing kinetic theory and
diagrammatics \cite{landaulifshitz},
we present a reaction-rate formula valid when the plasma is
slightly out of chemical equilibrium. In Section 4,
we calculate the gravitino regeneration rate,
applying the effective Lagrangian for
light gravitinos obtained in Section 2 and the
finite-temperature formalism given
in Section 3. In Section 4 the regeneration
rate is expressed as a discontinuity of the Goldstino
self-energy across an appropriate kinematical cut relevant
to the Goldstino regeneration process.
For a wide temperature range, including that of
phenomenological interest to us, we
find that the regeneration rate is qualitatively the same
as the estimate made at zero-temperature
 \cite{ekn,morth}. Compared to the zero-temperature
rate, the finite-temperature corrections contribute
$\sim \log(1/ \alpha_s(T))$, which results from finite-temperature
effects at temperature between $\sim g T$ and $ \sim T$.
Next-order corrections are proportional to
$\alpha_s(T) \, \log(1/\alpha_s(T))$,
mainly from thermal QCD corrections to
the Goldstino-matter coupling vertex. Therefore we conclude that
thermal effects do not alter drastically the temperature dependence
of the reaction rate as Fischler claimed. In Section 5,
we discuss the implications of our results
for cosmology focusing in particular on refined treatment of
reheating dynamics and implications to the baryogenesis, and in
Section 6 we summarize our results.

\section{Low-Energy Effective Lagrangian for Light Gravitinos}
We first recapitulate the relevant
effective Lagrangian for light gravitinos\cite{fayet},
in which supersymmetry is broken by the standard super-Higgs effect
in $N=1$ supergravity theory. Since the
gravitino interacts with gravitational strength, the
relevant terms in the $N=1$ supergravity Lagrangian are
always suppressed at least by one power of Planck scale
$M_P \equiv 1/ {\sqrt G_N}$.
We denote the chiral and gauge supermultiplets by
$(\phi, \chi)$ and $(A_\mu, \lambda)$. Keeping only the
leading dimension-5 terms, the
relevant Lagrangian for an on-shell gravitino in the
$\gamma^\mu \psi_\mu = 0$ gauge is given by
\be
{L \over {\sqrt -g}}= { 1 \over \sqrt 2 M_P}
[ D_\mu {\overline \phi}
{\overline \psi}_\nu \gamma^\mu \gamma^\nu \chi_R
+ D_\mu {\overline \phi}
{\overline \chi}_L \gamma^\nu \gamma^\mu \psi_\nu
+ { 1 \over \sqrt 2} {\overline \psi}_\mu
\sigma^{\nu \lambda}
\gamma^\mu \lambda F_{\nu \lambda}] + h.c.
\label{lagran}
\ee
When combined with a globally-supersymmetric
Lagrangian for matter and gauge multiplets,
the above Lagrangian may be used to calculate processes
involving the helicity-3/2 components of light gravitinos.

In addition to the helicity-3/2 components, the super-Higgs
effect from spontaneously-broken supersymmetry gives rise to
helicity-1/2 Goldstino components.
As long as the typical energy-momentum scale of the gravitino
is much larger than its mass $m_{3/2}$ and
$m_{3/2}$ is not larger than the effective mass splitting of
Bose and Fermi components for gauge and matter multiplets
( $m_{3/2}$ and the mass splittings may be decoupled as in the
case of no-scale supergravity \cite{noscale}), one can treat
the helicity-1/2 components of gravitinos as true Goldstinos.
This is the situation we study in what follows,
where we consider matter and gauge supermultiplet mass
splitting of order the weak scale, with a gravitino that
could be lighter.
More explicitly, in models in which supersymmetry is broken
by a hidden-sector field $S$ at a scale
$\langle S \rangle_{\theta^2} \equiv F \ne 0$,
and is coupled to the observable sector
at an intermediate-scale $M$,
the soft supersymmetry breaking mass splittings are generated by
the D term
\be
{1 \over M^2} \int d^2 \theta d^2 {\overline \theta^2} \,
S \, {\overline S} \, \Phi \, {\overline \Phi}
\label{scalarmass}
\ee
and by the F term
\be
{1 \over M} \int d^2 \theta \, S \, W_\alpha W^\alpha + h.c.
\label{gluinomass}
\ee
Therefore, mass splittings are of order
\be
\Delta m^2 \approx \left({|F| \over M }\right)^2  ;
\hskip1cm m_\lambda \approx {F \over M} .
\label{split}
\ee
For supersymmetry to be relevant to the stability
of the weak scale,
the soft supersymmetry-breaking mass splittings should be
of the same order as the weak scale $M_W$.
For simplicity, we denote both of these soft SUSY
breaking masses by a common mass $\ms$.
In this case, the gravitino mass is given by
\be
m_{3/2} = {|F| \over {\sqrt 6}M_P} \approx {M \over M_P}
\, m_{\rm soft}.
\ee
We consider $m_{3/2}$ no larger than $m_{\rm soft}$,
corresponding to $M$ no larger than $M_P$.

In the situation of interest to us, for which $m_{3/2}$ is much
smaller than the typical energy-momentum scale of the gravitino,
the low-energy theorem for the Goldstino implies
that the helicity-1/2 components of gravitinos are described by
\be
\psi_\mu =
{\sqrt {2 \over 3}} {1 \over m_{3/2}}  { i \partial_\mu} \psi
\label{longgrav}
\ee
where $\psi$ denotes the Goldstino and
$\sqrt {2/3}$ comes from the spin-1 Clebsch-Gordan coefficient.

Substitution of Eq.(\ref{longgrav}) into Eq.(\ref{lagran})
converts the dimension-5 operators of gravitinos into dimension-6
operators involving Goldstinos. These dimension-6 operators
in turn yield bad high-energy behavior of the Goldstinos compared
to the underlying supergravity theory.
As such the leading divergences of the helicity-1/2
gravitino-gravitino scattering amplitude present
in each individual diagram must cancel out in the total amplitude.
This is a consequence
of the fact that the helicity-1/2 components of
gravitinos are unphysical in the limit where
supersymmetry is unbroken and realized linearly. It is
desirable to rearrange the effective Lagrangian so that
this fact is manifest. In terms of Feynman diagrams this is done
by integrating by parts and using equations of motion for
external gauge and matter multiplet lines and subtracting the
leading divergence from each individual diagram for
internal gauge and multiplet lines. After this is done, the
interactions of the Goldstino with chiral and gauge
supermultiplets are given by
\be
L= - {\ms \over 8 F}
\bar \psi [\gamma_\mu, \gamma_\nu] \lambda \,
F^{\mu \nu} + {{\sqrt 2} \ms^2 \over F}
({\overline \chi}_R \psi) \phi + h.c.
\label{int}
\ee
The Lagrangian depends explicitly on the soft supersymmetry
breaking masses $\ms$, hence,
on the Goldstino-matter coupling constant $g_G = \ms^2 / |F|$.
We emphasize again that the Goldstino-matter coupling $g_G$
is nonzero only for softly-broken supersymmetry.

At finite temperature, two questions should be addressed
regarding the effective Lagrangian Eq.(\ref{int}).
First, the gravitino mass may receive thermal corrections.
If this correction is significant, one cannot apply the
low-energy theorem to evaluate the leading-order interactions
of gravitinos.
We have calculated the finite-temperature
correction to the gravitino mass using the interaction
vertices in Eq.(1). In the imaginary-time formalism of
thermal field theory, the correction is given in a form
$ \bar \psi_\mu \Sigma^{\mu \nu} (T) \psi_\nu$, where
$\Sigma^{\mu \nu} (T)$ denotes the two self-energy diagrams
at finite temperature depicted in Figure 1.
An explicit calculation based on the dimension-5 operators
in Eq.(\ref{lagran}) shows that the thermally-induced
gravitino mass is given by
\be
{1 \over 2} {\rm Tr}(\sigma_{\mu \nu} \Sigma^{\mu \nu}(T))
\approx {\pi^2 \over 12} \, {T^2 \over M_P^2} \ms.
\label{gravcorr}
\ee
The manifest chiral invariance present in unbroken
supergravity forbids a thermally-induced gravitino mass
for $\ms = 0$.
This underlies the dependence on the soft SUSY breaking mass
$\ms$ in Eq.(\ref{gravcorr}).
Comparing the result to $m_{3/2}$, we see that the thermal
correction is completely negligible as long as
$ T \ll \sqrt {M M_P} $. Even in models with $M \sim 10 \, TeV$,
the minimum temperature needed to obtain
an appreciable thermal mass correction is $\sim 10^{11} $ GeV.

Secondly, the Goldstino-matter coupling may receive
thermal corrections as depicted in Figure 2.
Recall that $g_G  = \ms^2 / |F|$  depends  explicitly on
the soft supersymmetry breaking masses $\ms$.
Therefore, thermal corrections to  $\ms$ and $F$ may in turn
induce thermal corrections to the coupling $g_G$.
One might expect that the soft supersymmetry breaking mass
$\ms^2$ receives a thermal correction of the form
$\ms^2 \rightarrow \ms^2 + {\cal O}(T^2)$.
This in turn might induce $\sim T^2$ thermal correction
to the coupling $g_G$ as in the Fischler's evaluation of the
reaction rate.  However, these steps are questionable.
First, even for the plasma heat bath,
the interaction of the gravitino with other matter
is described by a Noether coupling of the form
$\psi_\mu S^\mu + h.c.$.
Secondly, the soft masses $\ms$  are generated by the
interactions Eqs.(\ref{scalarmass}, \ref{gluinomass})
in the form of Eq.(\ref{split}).
This tells us that, for a fixed intermediate scale $M$,
thermal corrections to
$g_G$ from the Goldstino dynamics itself mainly originate
from thermal corrections to the supersymmetry breaking F-term
$ F = \langle S \rangle_{\theta^2}$.

There are actually two sources of thermal corrections
 to $g_G = \ms^2 / |F|$.
The first is the thermal mass correction to $\ms$ due to
 the Goldstino dynamics.
At the one loop level, the Goldstino interaction
Eq.(\ref{int}) yields
\be
\ms^2 (T)  \approx
\ms^2 \, (1 -{1 \over 24} {T^2 \over |F|} \,).
\ee
At the same order,  the F term receives a thermal correction
from the thermal Goldstino wave-function renormalization
\be
|F (T)| \approx |F| \, (1 - {1 \over 48} {T^2 \over |F|} \,).
\ee
Thermal tadpole corrections from interactions in
Eqs.(\ref{scalarmass},\ref{gluinomass}) only give rise to
$|\Delta F(T)| \approx T^2/M^2$, and hence are negligible
as long as $|F| \approx M \cdot M_W \ll M^2$.
Therefore, we find
thermal Goldstino dynamics induces a correction of size
\be
g_G (T)
\approx g_G \, (1 -{1 \over 48} {T^2 \over |F|} + \cdots)
\label{gcorr}
\ee
Again, the correction Eq.(\ref{gcorr}) is numerically
negligible for temperature of interest to us:
$T \ll \sqrt {|F|}$.
The actual thermal Goldstino correction might even be
further suppressed if
there is a direct extension of the finite-temperature
 nonrenormalization of the
Goldberger-Treiman relation \cite{eletsky} for the
 chiral dynamics of pions and
nucleons to the Goldstino dynamics of spontaneously-broken
supersymmetry.

Additional corrections to the
$g_G$ coupling come from the gauge interactions.
Unlike the thermal correction from the Goldstino dynamics,
 such corrections from
gauge interactions may be taken into account self-consistently
using the effective Lagrangian Eq.(\ref{int}) and
the resummed perturbation method.
Through explicit calculation in Section 4, we will see
that the thermal mass correction to $\ms$
only affects the thermal phase space but does not alter the
coupling $g_G$ from its zero-temperature value,
and that the thermal vertex correction is of higher order in
$\alpha_s(T) \log(1/\alpha_s(T))$.

The above result has an important consequence when we
consider the coupling of the helicity-1/2 Goldstino to the
supercurrent. Subtracting the leading divergent
contributions from each operator, we find that the
Goldstino coupling $g_G$ is essentially the zero-temperature
value, whereas possible thermal corrections from gauge
interactions are automatically taken into account by resummed
thermal propagators and vertices inside a loop diagram when
we calculate the regeneration rate.  Thus the Goldstino-matter
coupling $g_G$ ensures manifest decoupling of the Goldstino
from the rest of the theory in the limit where the
supersymmetry breaking in the observable sector
vanishes $\ms \rightarrow 0$.

Concluding this Section, for the range of supersymmetry
breaking scale of phenomenological interest,
we are justified in using the Goldstino
equivalence theorem and effective Lagrangian Eq.(\ref{int})
of the light gravitinos in the rest of this paper.

\section{Reaction Rate at Finite Temperature: Formalism}
We next formulate the particle reaction rate
in a plasma, using kinetic theory \cite{landaulifshitz}.
Consider a heat bath consisting initially
of thermal particles $q$'s and an isolated
particle species $\Phi$ slightly out of equilibrium.
The heat bath of $q$ particles at temperature $T$
may be described by a density matrix
\be
\rho_q=
{1 \over Z_o} \sum_{\{q\}} e^{-H/T} |q \rangle \langle q|.
\ee
The in-state density matrix of the heat bath plus
a single $\Phi$ particle is given by
\be
\rho_{in} \equiv |in \rangle \langle in | = a_\Phi^\dagger
\, \rho_q  \, a_\Phi
\ee
The $\Phi$ particles then interact with other particles
in the heat bath, and subsequently
decay into a new thermalized out-state density matrix
$\rho_{out}
\equiv |out\rangle \langle out|$. To be specific, we consider
two-to-two scattering of a $\Phi$ particle on $q$ particles
in the heat bath:
$\Phi(p_1) + q(p_2) \rightarrow q(p_3) + q(p_4)$,
whose reaction rate we denote by $\Gamma_f$.
For simplicity, we suppress
quantum numbers other
than the energy and momentum of each particle.
This thermal reaction differs
from zero-temperature decay, in that $\Gamma_f$ does not
specify completely the subsequent distribution of $\Phi$
particles. The inverse reaction $q(p_3) + a(p_4) \rightarrow
\Phi(p_1) + q(p_2)$,
whose reaction rate we denote by $\Gamma_b$,
is also possible in general,
and will create $\Phi$ particles out of the thermal bath.
Unitarity dictates that $\Gamma_f / \Gamma_b = \exp (E/ T)$,
irrespective of any possible violation of CP.
Let us denote the initial non-equilibrium distribution
function of the $\Phi$ particle by $n_i(E_1)$.
Subsequently, the $\Phi$ particle
distribution $n_\Phi(E_1)$ evolves according to the forward
and inverse reaction rates \cite{landaulifshitz,weldon}
\be
{d n_\Phi (t; E_1) \over d t} =
- n_\Phi (t, E_1)  \, \Gamma_f +
(1 \pm n_\Phi (t; E_1) ) \, \Gamma_b.
\label{rateeqn}
\ee
The $\pm$ signs for the inverse process are for
Bose-Einstein and Fermi-Dirac statistics of
the $\Phi$ particle, respectively.
Solving this equation, we find
\be
n_\Phi [t; E_1]
= n_{B,F} [E_1] + exp[-(\Gamma_f \mp \Gamma_b) t]
\label{ratesol}
\ee
where $n_{B,F}[E] = 1/(e^{E/T} \mp 1)$ for boson and fermion
equilibrium distribution functions, respectively.
Therefore, regardless of the initial distribution, the
primordial plasma approaches chemical equilibrium at
sufficiently late time.
As long as the starting assumption that the primordial
plasma is only slightly out of equilibrium is valid,
$\Gamma_f$ and $\Gamma_b$ can be calculated using the
equilibrium distributions $n_{B,F}[E]$. Under this
assumption, the rate of approach to equilibrium is
governed by $\Gamma_{tot} \equiv \Gamma_f \mp \Gamma_b$.

It now remains to calculate the total reaction rate
$\Gamma_{tot}$. The decay probability is given by
\bee
{\cal P} &=&
\sum_{out} \langle out | \Delta H \, \rho_{in} \,
\Delta H^\dagger | out \rangle \nonumber \\
         &=& {1 \over Z_o} {\rm Tr}
(e^{-H/T} \, a_\Phi \, \Delta H \, \Delta H^\dagger \,
a_\Phi^\dagger)
\label{prob}
\eee
where the last formula is given in terms of \sl retarded
\rm Green functions.
Denoting the thermal phase-space volume by $d \Omega$,
the reaction rate is given by
\bee
\Gamma_{tot} &=&  \int {1 \over 2 E_\Phi} {\cal P} d \Omega
\nonumber \\
       &=& {N_{pt} \over 2 E_1} \int\!\!\int\!\! \int \!
       {d^3 {\vec p}_2 \over (2 \pi)^3 2 E_2} \,
       {d^3 {\vec p}_3 \over (2 \pi)^3 2 E_3} \,
       {d^3 {\vec p}_4 \over (2 \pi)^3 2 E_4} \,
       n(p_2) \,(1 \pm n(p_3))\, (1 \pm n(p_4)) \nonumber \\
       &\times & (2 \pi)^4 \delta^{(4)}(p_1 + p_2 - p_3 - p_4)
       \, {\rm Tr}
  (  a_\Phi \, \Delta H \, \Delta H^\dagger \, a_\Phi^\dagger)
\label{rate}
\eee
Here $N_{pt} = (2s+1) \, N_c \, N_f$ denotes a factor of
summing over the spin, color, flavor etc. of the particle 2
in the initial heat bath.

The thermal correlation function is evaluated with
\sl retarded \rm boundary conditions.
As such, it is naturally interpreted as a discontinuity
across the physical cut relevant to the chosen process.
Due to the presence of the thermal bath,
there are new physical cuts present only at finite
temperature.  The discontinuity of the Green function is
related to the \sl imaginary part \rm of the self-energy
function of the $\Phi$ particle \cite{weldon}:
\be
\Gamma_{tot}  =
\Gamma_f \mp \Gamma_b = -{1 \over E } \, {\rm Im} \Pi [E +
i \epsilon]
\label{imaginary}
\ee
where $\Pi[E] = \Sigma[E]$ for bosonic $\Phi$ and
$\bar \Phi(E) \Sigma [E] \Phi(E)$ for fermionic $\Phi$.
In particular, the process we are interested in is given
by a cut which is present only for nonzero temperature.
The real part of the self-energy function is related to
the thermal correction to the mass of the particle.

\section{Thermal Reaction Rate for Light Gravitinos}
At the end of inflation,
the Universe is out of chemical equilibrium,
with the gravitino density depleted by the preceding
inflationary period.
However, scattering among particles in the primordial
plasma can regenerate gravitinos.
In this section,
we calculate this gravitino regeneration rate, considering
for definiteness the 2-to-2 scattering processes
$gluino + gluino \rightarrow gluino + Goldstino$ or
$gluon + gluino \rightarrow gluon + Goldstino$ \cite{ekn}.
Without taking finite-temperature effects into account,
it was shown previously
that $2 \rightarrow 2$ scattering was the dominant
mechanism of gravitino regeneration. We are interested
in the temperature range
$m_{3/2}, \ms <\!\!< T <\!\!< {\sqrt {|F|}}$.
Further, we assume that the regenerated
gravitinos are only slightly out of thermal equilibrium,
since it is only in this situation that we can apply
the equilibrium finite-temperature formalism discussed
in Section 3, and that $E$(gravitino),
$|\vec p|$(gravitino) $\sim {\cal O}(T)$,
so that helicity-1/2 gravitinos are the dominant
components for the scattering process.
This enables us to calculate the regeneration rate using
the Goldstino effective Lagrangian given in Section 2.

According to Eq.(\ref{imaginary}),
the regeneration rate of Goldstinos with an energy $E$,
a helicity $\lambda$ and an invariant mass-squared
$s \equiv k^\mu k_\mu$ is given in terms of
the discontinuity of the Goldstino self-energy
diagram $\SG$ depicted in Figure 3 by
\be
\Gamma_\lambda (E) = {i \over 2 E}\, { \rm disc} \,
({\overline u}_\lambda (k) \,
\Sigma_G (E, \vec k) \, u_\lambda (k))
\label{rate41}
\ee
where the Goldstino spinor wave function is denoted by
$u(k)$, and satisfies
\be
( \gamma^0 E - {\vec \gamma} \cdot {\vec k}) u (k) = 0
; \hskip1cm {\overline u}(k) u(k) = 2 {\sqrt s}.
\label{wavefunc}
\ee
We recall that the decay probability of the
Goldstino is related to the regeneration probability,
once the discontinuity in Eq.(\ref{rate41}) is taken across
an appropriate kinematical cut in the complex $E$ or $s$ plane.
However, the stimulated-emission factor $n_F(E)$ relevant for
the Goldstino decay rate should be replaced by
the Pauli-blocking factor $1- n_F(E)$
relevant for the regeneration rate.
Summing over the Goldstino helicity states $\lambda$,
we obtain the total regeneration rate
\bee
\Gamma_G (E) & = &
\, \, {i \over 2 E} \, (1 - n_F(E)) \, {\rm disc} \,
\sum_{\lambda} {\overline u}_\lambda (k) \,
\Sigma_G (E, {\vec k} ) \,
u_\lambda (k) \nonumber \\
& = & -{1 \over E} \, (1 - n_F(E)) \, {\rm Tr} \,
[ ( k \hskip-0.22cm / + {\sqrt s}) \,
{\rm Im} \Sigma_G (E + i \epsilon, {\vec k})] .
\label{totrate}
\eee
In what follows, we consider near-on-shell Goldstinos
and set $s = 0$. The Goldstino self-energy
$\Sigma_G (E, {\vec k})$ is calculated using the
effective Lagrangian of Section 2.
We emphasize again the important conclusion drawn in Section 2
that we can use the zero-temperature Goldstino-matter coupling
$g_G = \ms^2/|F|$ when evaluating $\SG$.
Finite-temperature corrections to $g_G$ will be taken
automatically into account by the resummed propagators
and vertices.
The imaginary part of $\Sigma$ can be expressed as a sum of
integrals over the phase space of initial and final
heat-bath states weighted by statistical
distributions. The integrands are squares of amplitudes
of the form (Goldstino)+2 $ \leftrightarrow$ 3+4,
where 2,3,4 are particles in the
plasma heat bath, which we have taken as gluinos and gluons.

In thermal field theory,
the Goldstino self-energy diagram may be calculated
using effective thermal vertices and propagators.
The skeleton diagram given in Fig.3 is then evaluated
in Minkowski space as ($k^\mu \equiv (E, {\vec k})$)
\be
\SG = \int {d^4 q \over (2 \pi)^4 }
\, q^\mu \Gamma_{\mu \alpha} (q) \, \Pi^{\alpha \beta} (q) \,
S_{\tilde g} (k-q) \, \Gamma_{\beta \nu} (q) \, q^\nu .
\label{selfenergy}
\ee
Here $\Pi^{\mu \nu}$, $S_{\tilde g}$ and $\Gamma_{\mu \nu}$
are effective thermal gluon and gluino propagators and the
effective thermal vertex for gluon-gluino-gluino coupling,
respectively, in which naive perturbative
diagrams are resummed. It is well known that in hot
gauge theories such a resummation is
necessary to take screening effects into account correctly.
So far, in gauge theories a consistent resummation method
has been developed only using the
imaginary-time formalism of thermal field theory,
mainly by Braaten and Pisarski \cite{braatenpisarski}.
The above regeneration rate formula Eq.(\ref{totrate}) has
the advantage that the Goldstino self-energy $\SG$ can be
calculated straightforwardly using the same imaginary-time
formalism.
We now apply the Braaten-Pisarski
method to calculate the Goldstino self-energy in a
resummed perturbative expansion.

{}From the structure of $\SG$ in Eq.(\ref{selfenergy}),
it is easy to see that the regeneration rate falls off rapidly
for a large momentum transfer to the internal gluon line, i.e.,
the $\vec q$ integrand has $ \sim 1/|\vec q|$.
This gives logarithmic divergences at small and large $|\vec q|$.
Therefore it is convenient to introduce some arbitrary scale
${\overline q}$ such that $gT <\!\!< {\overline q} <\!\!< T$ and
divide the  $|\vec q|$ kinematics into
soft  $g T \la |\vec q| <\!\!< {\overline q}$ and hard
$ {\overline q} <\!\! |\vec q| \la T$ regimes.
Since ${\overline q}$ is an arbitrary
scale, any physical quantity should be independent of it.

For the soft regime, the energy and momentum through the gluino
internal line are hard, namely
$E - q^0, |{\vec k} - {\vec q}| \sim {\cal O}(T)$, so
a bare gluino propagator suffices.
We only need to retain an effective gluon propagator which
 resums the bubble diagrams.
As we have argued earlier,
we also assume that the external Goldstino
has energy and momentum of order $\sim T$.
In this case a bare Goldstino-matter coupling $g_G = \ms^2/|F|$
suffices.
Only at higher orders or for Goldstinos
whose energy and momentum are less
than ${\cal O}(T)$ it is necessary take into account a different
effective Goldstino-matter coupling. We will discuss this later
in this Section.
For the hard regime,
both the gluon and gluino internal lines are hard.
Thus this contribution may be calculated
using bare vertices and propagators.
In the following calculation,
we take the leading logarithmic approximation,
for which the total reaction rate may be obtained just from
the calculation in the soft regime and the hard contribution
is automatically taken into account by the requirement
that the final result should be independent
of the arbitrarily introduced scale ${\overline q}$.

To cure the logarithmic divergence in the soft regime,
it is necessary to
resum the bubble diagrams of the gluon self-energy following the
Braaten-Pisarski method. As we will see, the resummation turns
the infrared-divergent contribution into an infrared-finite one.
It is known that the Braaten-Pisarski method is gauge-independent
for any physical observable such as the regeneration rate.
Therefore we are free to choose any convenient gauge,
and will choose Coulomb gauge in
the subsequent calculations. In this gauge,
denoting $q^\mu \equiv (q_0, {\vec q}), \,\, q \equiv |{\vec q}|$,
the nonzero components of the effective
gluon propagator $\Pi^{\mu \nu}$ after the resummation are known
\cite{effgluon}
\bee
\Pi^{00}(\omega, q) &=& \Delta_L (\omega, q),
\nonumber \\
\Pi^{ij}(\omega, q) &=&
\Big( \delta^{ij} - {q^i q^j \over q^2} \Big)\,
\Delta_T (\omega, q).
\nonumber \\
\label{gluonprop}
\eee
Here
\bee
\Delta_L (\omega, q) \, & \equiv & \,
\Big[ \, q^2 -{3 \over2} m_g^2 \,
\Big\{ {\omega \over q} \, \ln \Big({ \omega + q \over \omega - q}
\Big) - 2 \Big\} \Big]^{-1} ,
\nonumber \\
\Delta_T (\omega, q) \, & \equiv & \,
\Big[ \, \omega^2 -q^2 +{3 \over 2} \, m_g^2 \,
\Big\{ {\omega \over 2} \, \Big( {\omega^2 \over q^2} -1 \Big)
\, \ln \Big({\omega + q \over \omega - q} \Big)
- \Big({\omega \over q})^2 \Big\} \Big]^{-1}
\label{gluonprop2}
\eee
in which
$m_g^2 (T) \equiv {8 \pi \over 3}(1 + n_f /6) \alpha_s(T) T^2$
($n_f$ = number of light quark flavors in the
fundamental representation)
is derived for the supersymmetric particle content in the
high-temperature limit $\omega, q <\!\!< T$ and may be
interpreted as an effective gluon mass generated by interactions
with the primordial plasma.

Using this, the self-energy $\SG$ Eq.(\ref{selfenergy})
simplifies to
\be
\Sigma_G (E, {\vec k}) = i |{ \ms \over 8 F}|^2
\int {d^4 q \over (2 \pi)^4} \, q^\mu \gamma_{\mu \alpha} \,
{\Pi^{\alpha \beta}(q) \over
k \hskip-0.23cm / - q \hskip-0.225cm / - \ms} \,
\gamma_{\beta \nu} q^\nu .
\label{selfenergy1}
\ee
Next we take the gamma matrix trace of the product of $\SG$ with
the Goldstino projection operator given in Eq.(\ref{totrate}):
\be
{\rm Tr} \, [k \hskip-0.23cm / \, \Sigma_G (E, {\vec k})]
= - \, |{\ms \over 8 F}|^2 \, \sum_{q_0} \, \int
{d^3 {\vec q} \over (2 \pi)^3} \,
{A \, \Delta_L (q_0, q) + B \, \Delta_T (q_0, q) \over
(E - q_0)^2 -({\vec k} - {\vec q} )^2 - \ms^2}
\label{trace}
\ee
where $A$ and $B$ are calculated using the on-shell condition
$E^2 -{\vec k}^2 = 0$ as
\bee
A = 4 \, \Big[&-&2 ({\vec k} \cdot {\vec q})^2 +
({\vec k} \cdot {\vec q}) \, q^2 + 2 q^2 E^2
 -q_0 \, q^2 \, E \, \Big] ,
\nonumber \\
B = 8 \, \Big[ &-&
({\vec k} \cdot {\vec q}) \, (\, 3 {\vec k}
\cdot {\vec q} - q^2 \, ) +E^2 \, q^2
+ q_0 \, \Big\{ 4 {\vec k} \cdot {\vec q} - q^2 \Big\}\, E
\nonumber \\
&+& q_0^2 \Big\{ -3 E^2 -({\vec k} \cdot {\vec q})
+ ({\vec k} \cdot {\vec q})^2 /q^2 \, \Big\}
+ q_0^3 \, E \, \Big] .
\label{AB}
\eee
A possible soft gluino-mass dependence in the numerator drops
out after the trace over the gamma matrices is taken.
To simplify further calculations, we have also
ignored the soft gluino mass in the gluino propagator.
Next we sum over the discrete energies $q_0 = 2 \pi i n T$.
This is most easily done using the spectral representation for
the gluino and the effective gluon propagators.
For the gluino propagator
\bee
&{} & {1 \over (E - q_0)^2 -({\vec k} - {\vec q})^2}
= -{1 \over \pi} {\rm Im}
\Big[ \, {1 \over (E - q_0 + i \epsilon)^2
-({\vec k} - {\vec q})^2} \, \Big]
\nonumber \\
&=& \! -{1 \over 2 |{\vec k} - {\vec q}|}
\! \int_0^{1 \over T} \! d \tau_2 \,
e^{(E - q_0) \tau_2 } \Big\{ (1 - n_F(|{\vec k} - {\vec q}|)) \,
e^{- |{\vec k}-{\vec q}| \tau_2} - n_F(|{\vec k} - {\vec q}|) \,
e^{+|{\vec k} - {\vec q}| \tau_2} \Big\}.
\label{gluinospectral}
\eee
For the effective gluon propagator \cite{pisarskiphysica}
\bee
\Delta_L(q_0, q) &=&
	 -{1 \over \pi} {\rm Im} \Delta_L ( q_0 + i \epsilon, q)
= -\int_0^{1 \over T} d \tau_1 \, e^{ q_0 \tau_1}
\int_{-\infty}^\infty \! d \omega \,\,
		   \rho_L (w, q) \, e^{-\omega \tau_1} \,
[1 + n_B(\omega)]
,
\nonumber \\
\Delta_T(q_0, q) &=&
         -{1 \over \pi} {\rm Im} \Delta_T ( q_0 + i \epsilon, q)
= -\int_0^{1 \over T} d \tau_1 \, e^{q_0 \tau_1}
\int_{-\infty}^\infty \! d \omega \,\,
		    \rho_T (w, q) \, e^{-\omega \tau_1} \,
[1 + n_B(\omega)]
{}.
\label{gluonspectral}
\eee
They have nonzero support in the complex $\omega$ plane:
\bee
\rho_L (\omega, q) &=& {3 \over 2} m_g^2 {\omega \over q} \,
|\Delta_L (\omega + i \epsilon, q)|^2
\nonumber \\
&=& {3 \over 2} m_g^2 {\omega \over q} \,
\Big[ \, {9 \pi^2 \over 4} m_g^4 \Big({\omega \over q} \Big)^2
+ \Big\{\, q^2 + 3 m_g^2 -{3\over 2} m_g^2 {\omega \over q} \,
\log \Big( {q+ \omega \over q - \omega} \Big) \, \Big\}^2 \,
\Big]^{-1} ,
\nonumber \\
\rho_T (\omega, q) &=& {3 \over 4} m_g^2 \Big(\, {\omega \over q}
-\big( {\omega \over q} \Big)^3 \Big)
\, |\Delta_T (\omega + i \epsilon, q)|^2
\nonumber \\
&=& {3 \over 4} m_g^2 \Big(\, {\omega \over q}
-\Big( {\omega \over q} \Big)^3 \Big) \,
\Big[
{9 \pi^2 \over 16}
m_g^4 \Big( {q \over \omega} - {\omega \over q} \Big)^2
\nonumber \\
&+& \Big\{ \, q^2 - \omega^2 + {3 \over 2} m_g^2
\Big( {\omega \over q} \Big)^2 \,
\Big( 1 + {1 \over 2} \Big( {q \over \omega} - {\omega \over q}
\Big) \, \log \Big( {q + \omega \over q - \omega}\Big)
\, \Big) \Big\}^2 \, \Big]^{-1}
\label{support}
\eee
for $-q < \omega < + q$ in addition to the
plasmon and the transverse gluon delta function poles at
$\omega = \pm \omega_L (q), \pm \omega_T(q)$ respectively.

After rewriting the propagators in
terms of these spectral representations, it is easy to see that
the sum over discrete $q_0$ yields expressions of the form
\be
\sum_{q_0} \, q_0^n \, e^{q_0 (\tau_1 - \tau_2)} =
({\partial \over \partial \tau_1})^n \, \delta (\tau_1 - \tau_2)
\label{qsum}
\ee
for $n = $ 0, 1, 2, 3, relevant for $A$ and $B$ in Eq.(\ref{AB}).
Integration over $\tau_{1,2}$ then yields a factor of the form
\be
- {1 \over E \mp |{\vec k} - {\vec q}| - \omega} \,
\Big( \, e^{(\mp  |{\vec k} - {\vec q}| - \omega)/T}
+ 1 \Big)
\label{energydenom}
\ee
in which we have used the fact that
$E = 2 \pi i (n + 1/2) T$ for the Goldstino.
After all these steps are taken, the overall dependence of
${\rm Tr} \, [ k \hskip-0.23cm / {\rm Im} \SG]$
resides only in the energy denominator in Eq.(\ref{energydenom}):
\be
{\rm Im} {1 \over (E +  i \epsilon) \mp
                      |{\vec k} - {\vec q}| -\omega }
= - \pi \, \delta(E \mp  |{\vec k} - {\vec q}| - \omega)
\ee
Therefore, we finally obtain for
the imaginary part of the trace in Eq.(\ref{trace}):
\bee
{\rm Tr} &[& k \hskip-0.22cm / \, {\rm Im} \SG \, ]
 \nonumber \\
 &=& - \pi \, |{\ms \over 8 F}|^2 \, {1 \over (1 - n_F(E))}
 \nonumber \\
&\times& \int {d^3 {\vec q} \over (2 \pi)^3}
	    \int_{-\infty}^\infty d \omega \,
[{\cal A} \,( \rho_L (\omega, q) + {\cal B}\,\rho_T(\omega, q)]
\nonumber \\
&\times & {1 \over 2{\cal E}} \, (1 + n_B(\omega))\,
[\, (1 - n_F({\cal E})) \,
\delta (E - {\cal E} - \omega) - n_F ({\cal E}) \,
\delta (E + {\cal E} - \omega) \, ]
\label{discfinal}
\eee
in which ${\cal E} \equiv |{\vec k} - {\vec q}|$ and
$ {\cal A}, {\cal B} $ are given by
$A, B$ in Eq.(\ref{AB}) with $q_0$ replaced by $\omega$.

Recall that
$E, |{\vec k} - {\vec q}|, |{\vec k}| \approx {\cal O}(T) >\!\!>
q \approx {\cal O}(g T) > |\omega|$.
Because of this, the second delta function in Eq.(\ref{discfinal})
does not contribute at all.
In addition, $e^{-\omega/T} (1 + n_B(\omega))
\approx T/\omega$ and $|{\vec k} - {\vec q}| \approx k
+ {\vec k} \cdot {\vec q} / k$.
Then the angular $\vec q$ integration can be done straightforwardly
with the first delta function in Eq.(\ref{discfinal}) and
gives rise to a restriction to $|\omega| < q$,
unique to the finite-temperature regeneration process.
After these manipulations the reaction rate is given by
\bee
\Gamma_G (E) &=&  {2 \over \pi}  \, T \, |{\ms \over 8 F}|^2
\nonumber \\
& \times &
\int_0^{\overline q} q^3 \, dq \,
\int_{-q}^{+q} {d \omega \over \omega}
\, \Big\{ \, 1 - {\omega^2 \over q^2} \, \Big\}
\Big[ \, \rho_L(\omega, q) + \Big( 1 - {\omega^2 \over q^2} \Big)
\, \rho_T(\omega, q) \, \Big]
{}.
\label{rateintegral}
\eee
After appropriate rescaling and changing the order of integration,
we have done the $q$ integral analytically and the
$\omega$ integral numerically.
The result is given by
\be
\Gamma_G = {3 \over \pi} \, T\,  m_g^2(T) \,  |{\ms \over 8 F}|^2
\, \Big[
\, \log \Big( {{\overline q}^2 \over m_g^2} \Big)  - 1. 38
+ \cdots \Big]
\label{finalrate}
\ee
At small $\alpha_s(T)$, which is relevant to our case,
the first term is the dominant, leading-logarithmic contribution.
The second term is a subleading-logarithmic contribution.
As explained earlier,
the dependence on the arbitrarily-introduced scale $\overline q$
is a theoretical artifact.
Once we add the contribution from hard regime, this dependence
should disappear \cite{braatenyuan}.
Therefore, without an explicit calculation for the hard regime,
the regeneration rate in the leading-logarithmic approximation
can be obtained from Eq.(\ref{finalrate}).

In the above evaluation we have used the bare Goldstino-matter
coupling constant $g_G = \ms^2 /|F|$.
According to power counting, this is justified since the Goldstino
and the gluino have energy and momentum of order ${\cal O}(T)$.
The thermal-loop correction to the Goldstino vertex may be
evaluated by calculating the Figure 2 (a) and (b)
in which the blob consists of resummed,
effective internal propagators
on the gluon-gluino or chiral matter boson-fermion lines.
Let us consider Figure 2 (a). By power counting,
the vertex correction is quadratically divergent for small $q$
along the internal gluon line.
Therefore it is necessary to use the effective gluon propagator
Eq.(\ref{gluonprop}) as well as effective thermal masses for
the internal particles.
These resummations then soften the quadratic infrared divergence
of the integrand into a logarithmic one.
The result is that the Goldstino-matter coupling
receives a correction of the form
\be
g_G(T) = g_G \, ( \, 1 + \alpha_s(T) \,
\log(1 / \alpha_s(T)) + \cdots).
\label{vertexcorr}
\ee
The ellipses denote subleading ${\cal O}(\alpha_s(T))$
or higher-order corrections.

Combining Eq.(\ref{vertexcorr}) with the correction
Eq.(\ref{gcorr}) from the Goldstino dynamics,
we find that the gravitino regeneration rate is given by
\bee
\Gamma_G &=& {1 \over 16}\,
\Big(1 + {n_f \over 6} \Big) \, |{\ms \over F}|^2 \, T^3 \,
\alpha_s(T) \, \log(1 /\alpha_s (T)) \nonumber \\
&\times&
\Big[ \, 1 + \alpha_s(T) \, \log(1/\alpha_s(T))
+ {T^2 \over |F|} + \cdots \Big].
\label{llresult}
\eee
We have suppressed model-dependent numerical factors of order
unity in each of the correction terms.

Eq.(\ref{llresult}) is the main result of this paper.
It shows that the regeneration rate calculated in the
 high-temperature limit
and in the leading-logarithmic approximation is proportional
to $T^3$.
This temperature dependence agrees qualitatively with
the one obtained from the zero-temperature calculation\cite{ekn}.
When summed over the minimal supersymmetric
standard model particle content,
the numerical prefactor in Eq.(\ref{llresult}) is $\sim 0.291$.
In the earlier calculation of \cite{ekn} the Boltzmann equation
for the gravitino number density $n_{3/2}$ was used to deduce the
regeneration rate
\be
\Gamma_{3/2} \approx {1 \over n_{rad}} {d \over d t} n_{3/2}(T)
\approx \Sigma_{\rm tot} n_{\rm rad}(T).
\label{bolt}
\ee
where the expansion rate of the Universe and possible contributions
from heavy-particle decay are neglected in Eq.(\ref{bolt}),
and the zero-temperature total cross section is denoted by
$\Sigma_{\rm tot}$. From the result of \cite{ekn} it is easy
to check that the leading-logarithmic contribution is essentially
the same as ours in Eq.(\ref{llresult}), except for a
slightly different numerical prefactor $\sim 0.250$ for
the minimal supersymmetric model particle content
(after correcting discrepancies by
factors of 2 in the Goldstino effective Lagrangian used
in \cite{ekn}).
The finite-temperature calculation, however, was essential
in our case
to soften infrared divergences and yield the finite result of
Eq.(\ref{llresult}) in the leading-logarithmic approximation.
In contrast, in \cite{ekn} the infrared divergence
was cut off in an {\it ad hoc}
manner. Furthermore the factor $\log(1/\alpha_s(T))$ implies that
the thermal
QCD effects mainly originate over the range from the extreme soft
$q \sim g T$ to the extreme hard  $q \sim T$ regime.

\section{Cosmological Implications}

In the absence of inflation, there are clear limits on the
gravitino mass \cite{weinberg}.
A stable gravitino (e.g., if the gravitino were the LSP)
would contribute to  the present overall mass density an amount
\begin{equation}
\rho_{3/2} = m_{3/2} n_{3/2} \sim O(10^{-2}) m_{3/2} n_\gamma
\end{equation}
where $n_\gamma$ is the present density of photons. In order that
$\Omega_{3/2} h^2 \le {1 \over 4}$, where $h$ is the present
Hubble expansion rate of the Universe in units of
$100 {\rm km s}^{-1} {\rm Mpc}^{-1}$, one requires
$m_{3/2} < O(1)$ keV.
More massive gravitinos must be unstable.
Because the gravitino decay rate is suppressed,
$\Gamma_{3/2} \sim m_{3/2}^3/M_P^2$,
gravitinos come to dominate the energy density of the Universe
prior to their decay,
when $T_D \sim m_{3/2}^{5/3}/Y^{1/3}M_P^{2/3}$
(see, e.g., \cite{nos}),
where $Y = n_{3/2}/n_\gamma \sim 10^{-2}$
in the absence of inflation.
In cosmologies with inflation, $Y$ may be much smaller,
but in this case gravitinos may never dominate the energy density
of the Universe.
Subsequent to gravitino decay,
the Universe ``reheats" to a temperature, $T_R \sim
m_{3/2}^{3/2}/M_P^{1/2}$ and, for the Universe to recover prior
to nucleosynthesis, we must have $m_{3/2} \ga 20$ TeV.  However,
even in this case one must still be concerned about the
dilution of the baryon-to-entropy ratio \cite{eln}, which
would be by a factor
$\Delta = (T_R/T_D)^3 \sim Y (M_P/m_{3/2})^{1/2}$.
Dilution may not be a problem if the baryon-to-entropy ratio
is initially large.

Inflation modifies the above constraints on the gravitino
mass \cite{eln}. During inflation, the abundance of
gravitinos relative to photons is dramatically reduced,
as is the abundance of many other unwanted relics.
The problem with gravitinos, however, is that they are
regenerated as the Universe rethermalizes after inflation.
If gravitinos satisfy the noninflationary bounds,
then their reproduction after inflation is never a problem.
For gravitinos with mass of order 100 GeV,
dilution without over-regeneration will also solve the problem,
but there are several
factors one must contend with in order to be cosmologically safe.
Gravitino decay products can also upset
the successful predictions of Big Bang nucleosynthesis \cite{ekn},
 and decays into LSPs (if R-parity is conserved) can also yield
too large a mass density in the now-decoupled LSPs \cite{ehnos}.

Let us consider first the constraints imposed by regeneration.
Using
the rate of gravitino regeneration
 given by  Eq. (\ref{llresult}), the
abundance of gravitinos produced after inflation will be
$Y \sim \Gamma / H$, $H \simeq \sqrt{N} T^2/M_P$,
where $N$ is the number of degrees of freedom.
The most stringent of the Big Bang nucleosynthesis
constraints comes from the photoproduction of deuterium and
$^3$He \cite{ens} and yields the limit
\begin{equation}
Y < 3 \times 10^{-14} ({100 {\rm GeV} \over m_{3/2}})
\end{equation}
coresponding to a limit on the reheating
temperature after inflation of
\begin{equation}
T_R < 2.5 \times
10^{8} ({100 {\rm GeV} \over m_{3/2}}) {\rm GeV}
\label{best}
\end{equation}
A slightly stronger bound (by an order of magnitude in $T_R$)
was found in \cite{km}.
It is evident that if the goldstino regeneration
rate were larger by a factor of $T^2/m_{3/2}^2$, then the left
side of (\ref{best}) would become $T_R^3/m_{3/2}^2$,
and the upper limit
on $T_R$ would be about $10^4 GeV (m_{3/2}/100$GeV).

Given the severity of even the previous limits,
e.g., eq. (\ref{best}),
it is important that this limit be put into perspective with
what we expect from typical inflationary models, and with what
we require from the point of view of Big Bang baryogenesis.
Inflationary models \cite{infl} can in principle be described
by a single dimensionful parameter $\mu$ which is fixed
\cite{nos,eeno3,cdo2} by the magnitude
of the observed microwave background fluctuations \cite{cobe}:
$\mu^2/M_P^2 \simeq {\rm few} \times 10^{-8}$.
Because it is assumed that the inflaton potential is of
magnitude $\mu^4$, the inflaton mass $m_\eta$, the duration
of inflation described by the number of e-foldings
$H \tau$, and the reheating temperature $T_R$
are all given by $\mu$:
\begin{eqnarray}
m_\eta & \sim
\mu^2/M_P  & \sim {\rm few} \times 10^{11} {\rm GeV}
\nonumber \\
H \tau & \sim   8 \pi (M_P/\mu)^2 & \sim  10^9
\nonumber \\
 T_R & \sim
\mu^3/M_P^2 & \sim  10^8 {\rm GeV}
\end{eqnarray}

However, in \cite{eeno3} it was noted that
in fact the Universe is not immediately thermalized subsequent
to inflaton decays,
and the process of thermalization actually leads to a smaller
reheating temperature,
\begin{equation}
T_R \sim \alpha^2 \mu^3/M_P^2 \sim 10^5 GeV~,
\end{equation}
 where $\alpha^2 \sim 10^{-3}$ characterizes the strength of the
interactions leading to thermalization.
This low reheating temperature
is certainly safe with regards to the gravitino limit (41)
discussed above.  Even if there is a more efficient reheating
mechanism leading to a higher reheating temperature $T_R$, or
if there is significant non-thermal gravitino production before
thermalization \cite{kofboy}, we do not expect the equivalent
$T_R$ to lie above (42),
which is also compatible with our gravitino bound (41).

It is sometimes asserted that a low reheating
temperature, (one compatible with
the bounds from gravitino regeneration) is incompatible with
baryogenesis above the weak scale.
However, what is important is not the
value of $T_R$, but rather the inflaton mass $m_\eta$.
With $m_\eta > 10^{11}$ GeV,
there are several possibilities for baryogenesis
besides electroweak baryogenesis \cite{osch}.  Even the simple
 out-of-equilibrium decay of GUT Higgs bosons with masses
 $m_H < m_\eta$ can generate a sizeable asymmetry.
 As long as inflaton decay can produce
these Higgses, they will be present and
out of equilibrium at a temperature
$T \ll m_H$ \cite{nos,dl}.
In this context, the Affleck-Dine mechanism \cite{ad},
involving flat sfermion directions of the scalar potential
also works quite efficiently \cite{eeno3}.
There is also the interesting mechanism proposed by
Fukugita and Yanagida \cite{fy1},
which generates a lepton asymmetry by the decay of a heavy
right-handed neutrino. Nonperturbative
electroweak interactions associated with sphaleron transitions
then reprocess this lepton asymmetry into a baryon asymmetry.
All that is required is that the mass of the
right-handed Majorana neutrino be less than the inflaton mass
\cite{cdo2}.
This provides an upper bound on the right-handed neutrino mass,
of about $10^{12}$ GeV, which in turn implies that
left-handed neutrinos cannot be arbitrarily light, and
suggests that they are likely to be in the range of
astrophysical interest, as was discussed explicitly
in the context of flipped SU(5) in \cite{eno3}.
Clearly there is no
difficulty in generating a baryon asymmetry while at the same
time satisfying the constraints imposed by gravitino
regeneration.

Note that {\em if} the goldstino production rate was
as large as claimed in \cite{fischler},
then gravitino masses less than a few TeV \cite{ens}
would be excluded,
as the nucleosynthesis contraints would imply that
$T_R < 10^4$ GeV. Heavier gravitinos would not be subject
to the nucleosynthesis bound, as in this case
gravitinos would have decayed before nucleosynthesis,
and only the weaker bounds, which
and are consistent with
$T_R \sim 10^5$ GeV, apply.

Finally, we would like to touch
briefly upon a related problem, namely that
excess entropy production by the scalar fields often thought
to be associated with local supersymmetry breaking,
namely the Polonyi problem \cite{pol}.  The Polonyi problem
is in general more dangerous than the gravitino problem,
as it is not resolved by inflation.
During inflation, the scalars are in general driven to
field values which differ from the global minimum after
inflation.
There is, however the possibility that these scalars are
in fact quite massive \cite{ekon}.  In no-scale
supergravity models, the masses of these scalars (and
the gravitino as well) are not determined at the tree level,
despite the local breaking of supersymmetry.
If these masses are large, as in \cite{ekon},
there is no longer a problem with
either the scalars or the
gravitino \cite{enq}. So long as the masses of the scalars
are more than $\sim 10^{-8} M_P$,
there is no appreciable entropy production
\cite{eno2}, and if the scalar masses are larger than
$\sim 10^{-12} M_P$ they never dominate the energy density
\cite{eeno3}, and are therefore acceptable.
Below this mass, the scalars do dominate, and their decays
to LSP's can lead to an excessive value for the present mass
density \cite{myy}.

\section{Conclusion}
In this paper, we have investigated the
light gravitino regeneration rate in the early Universe with
particular emphasis on finite-temperature effects.
We first found that thermal corrections to the gravitino mass
are negligibly small. We thus conclude that the
helicity-1/2 components of the
gravitino dominate reaction processes in the primordial plasma.
Using Weldon's discontinuity rule \cite{weldon},
we then related the Goldstino regeneration rate to the
discontinuity of the Goldstino self energy across the
appropriate kinematical cut in the complex energy plane.
To evaluate the Goldstino self energy, we used the
low-energy effective Lagrangian for helicity-1/2 components.
This is justified by the fact that
the thermal correction to the gravitino mass is completely
negligible, and the temperature is in the range
$\ms <\!\!< T <\!\!< {\sqrt F}$.

We have also examined thermal correction to
the Goldstino-matter coupling $g_G = \ms^2 /|F|$.
Naively one may expect that $\ms^2$ is replaced by
thermal mass $\sim \alpha_s(T) \, T^2$.
By careful study, we have found that no such large
correction arises.
Because all the thermal corrections are to be calculated
using the zero temperature $g_G = \ms^2/|F|$,
decoupling of the supersymmetry breaking $\ms \rightarrow 0$
must be regular even at high temperature.
We have shown this through detailed calculations.
Both the Goldstino dynamics
and the QCD interactions give rise to corrections of order
$T^2/|F|$ and $\alpha_s(T) \, \log(1/\alpha_s(T))$ respectively.
Numerically they are negligible since
$M_W \sim \ms <\!\!< T <\!\!< {\sqrt {|F|}}$.

At least up to the order we have calculated,
the Braaten-Pisarski method we have utilized does not exhibit
manifest supersymmetry since power-counting
and kinematical consideration dictates that
an effective gluon propagator is essential
although a bare gluino propagator is sufficient.
In the effective Goldstino Lagrangian,
the local supersymmetry of the
underlying supergravity theory is realized nonlinearly
as a global supersymmetry among Goldstino, chiral matter
and gauge supermultiplets.
A full order calculation may retain the Ward identity of
the global supersymmetry.
However,
even without appealing to the supersymmetry Ward identity,
our final result has shown that an earlier estimate of
the regeneration rate based on zero-temperature calculations
remains qualitatively correct,
up to leading logarithmic factor, which arises only in the full
resummed finite-temperature perturbation method.
In particular, at leading logarithmic approximation, we have
found that the regeneration rate was proportional to $T^3$.
This is in disagreement with Fischler's estimate that
the rate goes as $T^5$ at high temperature.

We also have discussed the implications of our results
for baryogenesis, and argued that they are compatible
with many plausible scenarios including the Affleck-Dine
mechanism and right-handed neutrino decay.

As we were finishing our work, we received a related
preprint by Leigh and Rattazzi \cite{leigh}.
They restricted themselves to supergravity theories with a
hidden sector in which the supersymmetry is transmitted to the
observable sector at high energy, near the Planck scale.
Their argument uses the supersymmetry Ward identity to
justify naive dimensional analysis, whereas our
results are more explicit. Whilst the supersymmetry
Ward identity may play some role,
we have not explicitly relied on it in our actual calculations.
Nevertheless, our conclusion agrees with theirs where we
overlap.

\section*{Acknowledgements}
SJR thanks Andrei Linde for several helpful discussions,
and the hospitality of the Theory Group at CERN
where this work was initiated.
This work was supported in part by
U.S. NSF-KOSEF Bilateral Grant(SJR),
KRF Nondirected Research Grant 81500-1341 (SJR),
KOSEF Purpose-Oriented Grant 94-1400-04-01-3
and SRC Program (SJR) and Ministry of Education BSRI-94-2418
(SJR);
DOE grant DE-FG05-91-ER-40633DOE (DVN);
DOE grant DE-FG02-94ER40823
 (KAO).

\begin{figure}

\caption{
Thermal mass correction to the gravitino mass. Crosses
denote insertions of the soft masses for the matter and
gluino fields. The dashed line represents a matter scalar field,
the solid lines matter fermion and gluino fields.}
\end{figure}
\begin{figure}
\caption{Thermal Goldstino vertex correction.
The blobs denote thermal corrections to the
gluino-gluon-Goldstino and to the matter
fermion-scalar-Goldstino couplings.}
\end{figure}
\begin{figure}
\caption{
Thermal correction to the Goldstino self-energy.
The blobs in the propagators are thermal self-energies
for the gluon and gluino respectively.
The blob is at the gluino-gluon-Goldstino vertex,
which is the same as
the zero-temperature vertex, as argued in Section 2.}
\end{figure}
\begin{figure}
\caption{
Two contributions to the Goldstino self energy.
(a) For hard thermal loops, the
resummed gluon propagator is the dominant effect.
(b) For soft thermal loops,
thermal corrections to the gluino and to the vertex
are the dominant effects.}
\end{figure}
\end{document}